\documentclass[useAMS,usenatbib]{mn2e}
\usepackage{graphicx,fleqn,times}

\title[Probing dark matter with X--ray binaries]
{Probing dark matter with X--ray binaries}

\author[W.~Dehnen, A.~R.~King] {Walter~Dehnen and Andrew~King\\
Theoretical Astrophysics Group, Department of Physics \& Astronomy, 
University of Leicester, Leicester LE1 7RH}

\date{Accepted 2005 December 05.
Received 2005 December 02; in original form 2005 November 16}

\volume{000}

\pagerange{\pageref{firstpage}--\pageref{lastpage}}
\pubyear{2005}

\begin{document}

\newcommand{\lsun} {{\mathrm{L}_\odot}}
\newcommand{\msun} {{\mathrm{M}_\odot}}
\newcommand{\rsun} {{\mathrm{R}_\odot}}
\newcommand{\mlsun} {{(M/L_V)_\odot}}

\maketitle
\label{firstpage}

\begin{abstract}
  Low-mass X-ray binaries (LMXBs), which occur in old stellar populations,
  have velocities exceeding those of their parent distribution by at
  least 20\,km\,s$^{-1}$. This makes them ideal probes for dark matter, in
  particular in dwarf spheroidals (dSph), where the LMXBs should penetrate
  well outside the visible galaxy. We argue that the most likely
  explanation of the observation of LMXBs in the Sculptor dSph by
  \cite{MaccaroneEtal2005} is the presence of a dark matter halo of $\ga
  10^9\,\msun$, corresponding to a total-mass to light ratio of
  $\ga600\mlsun$. In this case there should be an extended halo of LMXBs
  which may be observable.
\end{abstract}


\section{Introduction}
A recent study by \cite{MaccaroneEtal2005} reports the detection of five
X--ray binaries with $L_X > 6\times 10^{33}$~erg s$^{-1}$ in the Sculptor
dwarf spheroidal (dSph) galaxy. Their membership of Sculptor is
secure, as they are optically identified with counterparts in the
catalogue of \cite{SchweitzerEtal1995} which have appropriate proper
motions and photometry. Given the old stellar population of
Sculptor (see below) these must be low--mass X--ray binaries (LMXBs). In
the Milky Way, LMXBs are observed to have relatively high space velocities
$v_{\rm sp} \sim 20-100\,$km\,s$^{-1}$ (e.g.
\citealt{BrandtPodsiadlowski1995} and references therein; see
\citealt{PodsiadlowskiPfahlRappaport2005} for a recent review) resulting
from the recoil of the system after the supernova creating the compact
component. These velocities are far above the stellar velocity dispersion
$\sim11$\,km\,s$^{-1}$ in Sculptor, so it is interesting to ask how
Sculptor can retain any LMXBs at all.

\section{The nature of the LMXBs in Sculptor}
The observed LMXBs in Sculptor are relatively faint, and are probably
quiescent transients. Given a visible mass\footnote{Assuming a
mass-to-light ratio of $\sim6\mlsun$ and using $L_V=1.5\times10^6\lsun$ for
Sculptor, see Section~\ref{sec:dark:scl} below.} $\sim 10^7\msun$ for
Sculptor, the incidence $\sim 5 \times 10^{-7}\msun^{-1}$ of observed LMXBs
per stellar mass is significantly higher than that ($10^{-8} -
10^{-9}~\msun^{-1}$) of much brighter LMXBs in the large sample of
elliptical galaxies studied by
\cite{Gilfanov2004}.  This agrees with the deduction of
\cite{PiroBildsten2002} that the bright X--ray population of early--type
galaxies probably consists of the long--lasting outbursting transients
predicted by simple theories of the accretion disc outburst
(\citealt{KingKolbBurderi1996}, see also \citealt{RitterKing2002}).

The argument is straightforward: the low--mass donors ($\la 1\msun$) in
LMXBs cannot sustain the mass accretion rates $\sim 10^{-9} -
10^{-8}\msun$\, yr$^{-1}$ needed to power the bright sources in early--type
galaxies for a significant fraction of their ages, so accretion must be
intermittent. The relative incidence of faint X--ray sources in Sculptor
and of bright ones in ellipticals gives a lower limit to the ratio of
quiescent to outburst timescales (= 1/duty cycle) as $\ga 200$. This would
allow even the brightest sources in ellipticals to have ages comparable
with their host galaxy. Moreover it agrees with the completely independent
estimate of the duty cycle in long--period transients by
\cite{RitterKing2002}, who compared the observed numbers of long--period
neutron--star transients with their descendants (wide binaries containing a
millisecond pulsar and a low--mass white dwarf). This agreement suggests
that Sculptor has retained most of its X--ray binaries. We now ask how.

We can discard two possibilities. First, there is little reason to
suppose that supernova kicks have vastly different properties in
Sculptor and the Milky Way. Although the metallicity is low (see
below) the gross properties of LMXB kicks are not sensitive to such
details: kick velocities are of order the orbital velocities in the
pre--supernova binary because a symmetrical supernova explosion
carries off a mass comparable to that of the remaining binary with the
velocity of the exploding component. Anisotropies are important in
keeping bound some systems which would otherwise unbind, but do not
significantly alter this fact. The only way to increase the retention
rate on these lines is to remove the supernova explosion altogether,
i.e. to argue that the Sculptor LMXBs are all black--hole systems
which went through non-explosive collapses instead of supernovae. However
unless there is a special mechanism very strongly favouring
black--hole versus neutron--star formation we would require a very
high formation rate for LMXBs. Even then, black--hole LMXBs in the
Milky Way do appear to have kicks of 20 km\, s$^{-1}$ or more
\citep{TaurisVandenHeuvel2003} which would make them hard to retain in
Sculptor. In line with these arguments, low metallicities appear if
anything to lower LMXB incidence: \cite{KunduMaccaroneZepf2002} and
\cite{Bregman2006} show that globular cluster with lower metallicities
have fewer LMXBs. Rather than a systematic difference in kick
velocities, this effect probably indicates that metal--rich clusters
have higher LMXB tidal capture formation rates.

Second, it is implausible to argue that the observed LMXBs are not
bound to Sculptor: they would escape from a region of size $R =
R_{\mathrm kpc}$~kpc on a timescale $t _{\rm esc} \sim
R/v_{\mathrm{kick}} \la 2\times 10^7R_{\mathrm kpc}(50~{\rm
km~s^{-1}}/v_{\mathrm{kick}})\,$yr after the supernova, in many cases
not even evolving into contact and turning on as LMXBs until well
outside the galaxy. The required LMXB formation rate per unit galaxy
mass $\sim 5/(t_{\rm esc}M_{\rm gal}) \sim 2.5\times
10^{-15}\,$yr$^{-1}\, \msun^{-1}$ would be higher than inferred for
the Milky Way. There, with an LMXB population of $\sim 100/d$ and LMXB
lifetimes $\sim 1\,$Gyr one gets $10^{-18}d^{-1} \,$yr$^{-1}\,
\msun^{-1}$), where $d$ is the transient duty cycle. The very small
values $d \la 10^{-3}$ needed to make this agree with Sculptor would
create problems \citep[cf.][]{RitterKing2002} in spinning up neutron
stars to millisecond periods in LMXB descendant binaries, because the
neutron stars would accrete very little mass during the rare but
generally highly super--Eddington outbursts. We note further that the
pre--supernova lifetime of any star massive enough to make a neutron
star is (conservatively) $\la 10^8\,$yr, so any burst of star
formation making an overabundance of LMXBs would have to have been
more recent than that, which is ruled out by observation (see below).

A rather more promising idea, but still with difficulties, is that
there might exist a class of LMXB with low--velocity kicks which
Sculptor could retain, beyond the minority of black--hole systems
referred to above. \cite{PodsiadlowskiPfahlRappaport2005} indeed
identify a low--velocity subset of high--mass X--ray binaries, but
these can no longer be present in Sculptor's old stellar population.
Apart from the minority of black--hole systems discussed above, there
appear to be very few LMXBs with low--velocity kicks. Thus if such a
class exists it is clearly rather small, and one would again face the
problem of a high required birthrate of LMXBs. Worse, one would
require a still smaller transient duty cycle ($ d \la 10^{-3}$) to
match the incidence of LMXBs in Sculptor with that in ellipticals,
running into the same problem in spinning up neutron stars to
millisecond periods noted above.

These attempted explanations all tried to ascribe reduced space
velocities to the Sculptor LMXBs in order to compensate for the
galaxy's feeble apparent gravity. There appears to be only one fairly
straightforward explanation for the Sculptor LMXB population observed
by \cite{MaccaroneEtal2005}. The galaxy must have much stronger
gravity, i.e.\ dark matter.

\section{Dark matter in dwarf spheroidals} 
In recent years evidence has grown for the existence of extended massive
dark-matter haloes around some of the dSph satellites of the Milky
Way. Extensive radial--velocity data, for example for the Draco dSph,
suggest these objects are embedded in enormous dark--matter haloes
\citep{KleynaEtal2001}, resulting in total-mass to light ratios well in
excess of 100$\,\mlsun$. Moreover, such extended dark--matter haloes make
the star--formation process more comprehensible
\citep*{MashchenkoCouchmanSills2005} and alleviate the ``missing
satellites'' problem of cold dark matter cosmologies.

\subsection{The Sculptor dwarf spheroidal} \label{sec:dark:scl}
Sculptor appears to be the only Galactic dSph with no young stellar
population (that was why \citeauthor{MaccaroneEtal2005} chose it to search
for LMXBs). Its stellar population is as old as that of globular clusters,
but with a more extended star formation period, although a small tail of
residual star formation until about 2 Gyr ago cannot be ruled out
\citep{MonkiewiczEtal1999}.

Sculptor's stellar population contains two kinematically and spatially
distinct components of different metallicity, also reflected in a
bi-modality of its horizontal branch \citep[e.g.][]{TolstoyEtal2004}.
The ``metal rich'' ($\mathrm{[Fe/H]}>-1.7$) component is more centrally
concentrated than the metal poor component ($\mathrm{[Fe/H]}<-1.7$).
The line--of--sight velocity dispersion $\sigma$ of the former is
$\approx6\,$km\,s$^{-1}$ (\citealt{ArmandroffDaCosta1986};
\citealt*{QuelozDubathPasquini1995}; \citealt{TolstoyEtal2004}), while
the latter has $\sigma\approx11\,$km\,s$^{-1}$ \citep{TolstoyEtal2004,
  ClementiniEtal2005}.

From their observation of $\sigma\approx6\,$km\,s$^{-1}$,
\cite{QuelozDubathPasquini1995} derived a central mass--to--light ratio
$M/L$ of $13\pm6\,\mlsun$ and a total mass (within the visible galaxy) of
$(1.4\pm0.6)\times10^7\,\msun$ (assuming a single stellar component). The
same authors report an absolute magnitude of $M_V=-10.7\pm0.5$
corresponding to $L_V=(1.5\pm0.7)\times10^6\lsun$ and resulting in a
mass-to-light ratio of $10\pm6$, which is perfectly consistent with an old
stellar population, suggesting that dark matter is in fact
\emph{not} significantly contributing in the central region.

Although there has been no detailed modelling of the implications of the
bimodal stellar population with different $\sigma$, we may roughly assess
the consequences by noting that the mass--to--light ratio is proportional
to $\sigma^2$. Thus, from these data we may expect $M/L$ of up to
$\sim50\,\mlsun$, corresponding to a mass of $\sim6\times10^7\msun$ within
$\sim1.5\,$kpc (the apparent extent of the stellar population).

However, if the matter distribution of Sculptor did not extend beyond this
radius, one would expect the Galactic tidal field to distort Sculptor's
outer regions significantly, which is not observed
\citep*{ColemanDaCostaBlandHawthorn2005}. This lack of tidal distortion may
therefore be interpreted as indirect evidence for the presence of a more
extended dark--matter halo, like the ones claimed for Draco and Ursa Minor,
which protects the dSph from Galactic tides. The very fact that Sculptor
possesses two distinct stellar populations, evidently formed in two
starbursts, also strongly suggests that it is surrounded by an extended
dark--matter halo. This is needed to prevent the loss of all the gas expelled
during the first starburst in a galactic wind.

\subsection{LMXBs as probes of dark matter}
The binary kick velocity is typically $20-100\,$km\,s$^{-1}$, well in
excess of the velocity dispersion found in a dSph. The LMXBs of a dSph
should therefore be kicked out of the visible galaxy. If the dSph are
surrounded by an extended dark--matter halo, some of the LMXBs would
still remain bound, orbiting on highly eccentric, almost plunging,
orbits.  Otherwise most would escape and we would only see those
which received a rather small velocity kick. Thus, an extended
dark--matter halo for dSph galaxies implies a similarly extended halo
of LMXBs, a prediction that can be observationally tested. Since the LMXBs
would spend comparatively little time in the visible galaxy, there should be a
large number of them further out. The detectability of this halo obviously
depends on the surface density of the LMXBs, and thus on the kick velocities
and the dark matter distribution.

Because the LMXBs should form a distribution much more extended than their
parent population, they are ideal probes for more detailed investigations
of dSph dark--matter haloes. In particular, their velocities (radial and
proper motions) may be measured enabling in situ quantitative modelling of
the dark--matter haloes.

\subsection{LMXBs in Sculptor}
We now try to estimate roughly what kind of dark--matter halo is
required to keep the LMXBs observed in Sculptor bound. We model the
halo as a sphere with cumulative mass
\[
GM(<r) = v_0^2 \frac{r^3}{r^2+r_{\mathrm{c}}^2},
\]
which corresponds to a density profile similar to the
pseudo--isothermal sphere. Here, $v_0$ is the (asymptotic) circular
speed and $r_{\mathrm{c}}$ the core radius. We assume that the dark
mass within the visible galaxy is $5\times10^7\msun$, in agreement
with our above estimate for the mass--to--light ratio, requiring
\[
v_0 = 12\,\mathrm{km}\,\mathrm{s}^{-1}
\sqrt{1+\left(\frac{r_{\mathrm{c}}}{1.5\mathrm{kpc}}\right)^2}.
\]
If the halo is truncated at radius $r_{\mathrm{t}}$, then the velocity
$v_{\mathrm{esc}}(r)$ required at radius $r<r_{\mathrm{t}}$ to escape to radius
$r_{\mathrm{t}}$ (where the LMXB would be stripped from the dark--matter
halo by the Galactic tidal field) is
\[
v_{\mathrm{esc}}^2(r) = v_0^2 \ln
\frac{r_{\mathrm{t}}^2+r_{\mathrm{c}}^2}{r^2+r_{\mathrm{c}}^2}.
\]
\begin{figure}
  \centerline{\hfil \resizebox{85mm}{!}{\includegraphics{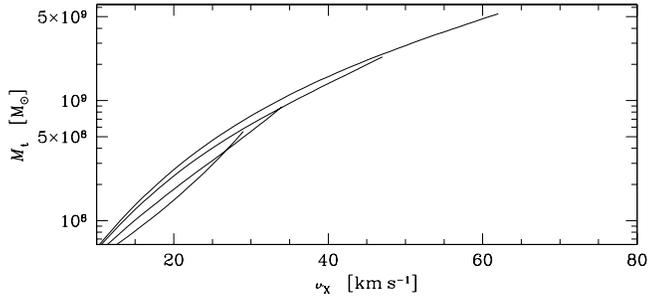}} } \caption{
  \label{fig:Mhalo} The minimum mass of an extended dark--matter halo for
  Sculptor that is required to hold on to an LMXB with velocity
  $v_{\mathrm{X}}$ at radius 1\,kpc for haloes with core radii
  $r_{\mathrm{c}}$ of 0.7, 1.5, 3, and 5\,kpc. The haloes are assumed to
  have mass $5\times10^7\msun$ within 1.5\,kpc. The curves end to the right
  when the required truncation radius $r_{\mathrm{t}}$ exceeds 15\,kpc (the
  larger the core radius, the larger the maximum $v_X$ still bound).  }
\end{figure}
In Figure~\ref{fig:Mhalo} we plot the corresponding halo mass required to
hold on to an LMXB moving out from $r=1\,$kpc with velocity $v_{\mathrm{X}}$
for various choices of $r_{\mathrm{c}}$. Evidently, this figure suggests
that it would be difficult for a dSph to hold on to LMXBs with velocities of
50\,km\,s$^{-1}$ or higher unless it has a total mass in excess of
$10^9\,\msun$ (this is largely independent of details, such as the core
radius of the dark matter or the initial position of the LMXB at the time
of the kick). Since $10^9\,\msun$ is about the dark mass discussed for
dwarf spheroidals \citep[e.g.][]{MashchenkoCouchmanSills2005}, we would
thus expect these galaxies to hold on only to the LMXBs with lower
velocities.  This argument can, of course, be turned around to provide a
mass estimator for the dSph dark--matter haloes via the escape--velocity
argument or more sophisticated dynamical modelling, usually applied to the
Milky Way.

\section{Discussion}
We have argued that the most likely explanation for the presence of LMXBs
in Sculptor is a dark matter halo of $\ga 10^9\,\msun$. This is $\ga 100$
times the observed stellar mass and corresponds to a total-mass to light
ratio of $\ga600\mlsun$. This value is comparable to that proposed for the
Draco dSph as a result of simulations of tidal stripping by
\cite{ReadEtal2005, ReadEtal2006}. We argued further that in this case there
should be an extended halo of LMXBs which might be observable. An estimate
of the total number of LMXBs in this halo would constrain the transient
duty cycle still further.

The obvious test of our ideas would come from measuring at least the 
radial velocities of the observed LMXBs. This should be possible given 
that \cite{MaccaroneEtal2005} were able to make optical identifications of 
some of these sources, and would be extremely interesting whatever the 
result. If the deduced space velocities are high, this would confirm the 
presence of a very massive dark matter halo. If not, this would have 
important implications for understanding transient duty cycles and LMXB 
formation in general.

\section*{Acknowledgements} 
ARK acknowledges a Royal Society--Wolfson Research Merit Award. Research in
theoretical astrophysics at the University of Leicester is supported by a
PPARC rolling grant.

\label{lastpage}
\end{document}